\documentclass[aps,prd,reprint,onecolumn,nofootinbib,amsmath,amssymb,superscriptaddress]{revtex4-2}
\usepackage[T1]{fontenc}
\usepackage{bm}
\usepackage[colorlinks=true,linkcolor=blue,citecolor=blue,urlcolor=blue]{hyperref}

\newcommand{\eps}{\bm{\epsilon}}
\newcommand{\heps}{\hat{\bm{\epsilon}}}

\begin{document}
\title{Thermodynamics of Kerr-Newman-Bertotti-Robinson black holes\\
from background subtraction}
\author{Chang-Xu Yan}
\email{cxyan@mail.bnu.edu.cn}
\affiliation{School of Physics and Astronomy, Beijing Normal University, Beijing 100875, China}

\author{Zili Mou}
\email{zilimou@bnu.edu.cn}
\affiliation{School of Physics and Astronomy, Beijing Normal University, Beijing 100875, China}

\author{Yu Tian}
\email{ytian@ucas.ac.cn}
\affiliation{School of Physical Sciences, University of Chinese Academy of Sciences, Beijing 100049, China}

\author{Hongbao Zhang}
\email{hongbaozhang@bnu.edu.cn}
\affiliation{School of Physics and Astronomy, Beijing Normal University, Beijing 100875, China}
\affiliation{Key Laboratory of Multiscale Spin Physics, Ministry of Education, Beijing Normal University, Beijing 100875, China}

\begin{abstract}

The non-asymptotic flatness of the charged rotating black holes immersed in the Bertotti-Robinson magnetic field causes the ambiguity in the gauge choice for the time generator and the electric potential. By using the newly established form of the integrability condition together with the gauge choice for the asymptotically flat charged rotating black holes as the initial condition, we have found that the gauge can be uniquely fixed. With this fixed gauge, we further obtain the mass and angular momentum by the background subtraction method with the Minkowski spacetime as the reference spacetime. As a result, not only is the first law of black hole thermodynamics as well as the quantum statistical relation automatically satisfied, but also both the Christodoulou-Ruffini relation and the Smarr formula are obeyed.

\end{abstract}
\maketitle

\section{Introduction}\label{sec:intro}

Due to Bekenstein’s insightful observation that black holes should be assigned an entropy proportional to the horizon area \cite{Bekenstein} and Hawking’s seminal discovery that black holes do radiate thermally with the temperature proportional to the surface gravity \cite{Hawking}, the four laws of black hole mechanics found originally for the Kerr-Newman (KN) black holes in \cite{Bardeen} are promoted as the four laws of black hole thermodynamics. Such a thermal nature of black holes, arising essentially from the quantum effect, is further substantiated by the Euclidean approach to quantum gravity \cite{Hartle,Gibbons}. In addition, it is proved by using the covariant phase space formalism in \cite{A24,A25} that the first law of black hole thermodynamics holds universally for general stationary black holes with some conventional asymptotic structures such as the asymptotic flatness and asymptotic AdS geometry, whereby not only can one choose a natural time generator to define the Arnowitt-Deser-Misner (ADM) mass at infinity but also fix the gauge for the electric potential by requiring it vanish at infinity. 

Among others, a new class of exact Type D stationary black hole solutions to the Einstein-Maxwell equations, called the Kerr-Newman-Bertotti-Robinson (KNBR) family, has recently  been obtained and analyzed in \cite{B22,B23,KNBR}. Indicated by its name, the KNBR black hole describes the charged rotating black hole immersed in the magnetic field of the uniform Bertotti-Robinson universe. 
The moduli space for the KNBR family is characterized by the four parameters $(m,a,e,B)$, related individually to the mass, angular momentum, electric charge, and the external magnetic field. In the absence of the magnetic field, namely in the case of $B=0$, the KNBR black hole reduces to the KN black hole. While in the presence of the magnetic field, the KNBR black hole is not asymptotically flat, which presents a challenge for one to explore its thermodynamics. For instance, in order to investigate the thermodynamic properties of the KNBR black hole, the authors of \cite{Li,Kubiznak} as well as \cite{ZZC} defined the mass by taking the Christodoulou-Ruffini (CR) relation \cite{Christodoulou} as an a priori assumption. Such a strategy is apparently ad hoc. So it is desirable to develop a systematic framework to address this issue. 

Actually there have been various advances made along these lines, all of which are based on the aforementioned covariant phase space formalism.
For one thing, the universality of the first law is generalized to the quasi-locally stationary isolated horizon, on which all involved quantities, including the mass, are evaluated without reference to the infinity and its associated asymptotic structure \cite{Ashtekar2000,Ashtekar2001}. 
But in this case, the requirement to have a consistent Hamiltonian evolution, which amounts to the so called integrability condition, leaves the freedom in choosing the time generator to define the horizon mass and gauging the horizon electric potential. Such a freedom can be fixed uniquely in an affirmative manner by further requiring that the aforementioned quasi-local quantities reduce to those obtained in the stationary KN black holes, where the event horizon serves as the isolated horizon. For another, the first law is also extended to some stationary black holes with unconventional asymptotics, where the uniqueness of the gauge choice is determined similarly by the integrability condition presented in \cite{Barnnich2002,Barnnich2003,Barnnich2008} together with the requirement that the result should recover that for the stationary black holes with the conventional asymptotics \cite{Barnnich2005,Barnnich2006,Compere2007,Astorino2016}. On the other hand, a correct criterion for the applicability of the background subtraction method has recently been established in \cite{RefA}, where not only is it found that the resultant criterion does not necessitate the isometric embedding requirement of the induced metric of the targeted spacetime into the reference spacetime, but also it is shown that such a criterion is met by both asymptotically flat and asymptotically AdS spacetimes for Einstein's gravity and its higher derivative corrections, correcting the prevailing misconception. 
It is noteworthy that this newly established criterion for the background subtraction method is essentially equivalent to the above integrability condition presented in \cite{Barnnich2002,Barnnich2003,Barnnich2008}. Both of them do not depend on the spatial radial position where the computations are performed. Namely, one can perform the relevant computation directly at the event horizon as done for the isolated horizon, or any other location outside of the black hole, for instance in the vicinity of the black hole or at infinity. But nevertheless, the former turns out to be much more amenable to analysis than the latter. 

With this in mind, we intend to apply the background subtraction method to investigate the KNBR black hole thermodynamics. To this end, we shall adapt the newly established criterion for the background subtraction method in the covariant phase space formalism to the Einstein-Maxwell theory in the subsequent section, where we also articulate its relation with the integrability condition. After a brief review of the KNBR black holes and the corresponding properties as well as the associated gauge freedom for the time generator and the electric potential in Section \ref{sec:KN-BR}, we shall first fix the gauge and then develop the KNBR black hole thermodynamics by using the background subtraction method in Section \ref{sec:subtraction}. We conclude our paper with some discussions in the final section. We relegate some straightforward but a little bit tedious calculations into Appendix \ref{app:boundary} and Appendix \ref{app:moduli-integrability}.

\textbf{Note Added}: While finalizing our draft, we became aware that \cite{Li2026} appeared on arXiv. Although the part of the conclusion overlaps, the formalism and the methodology we are using are apparently distinct from those used in \cite{Li2026}. In particular, the strategy we have developed is obviously innovative and the procedure we have presented is also much more transparent.

\section{Criterion for background subtraction method in Einstein-Maxwell theory }\label{sec:formalism}

We consider the Einstein-Maxwell theory with the following bulk Lagrangian form
\begin{equation}
    \mathbf L=\frac{1}{16\pi}(R-F_{ab}F^{ab})\boldsymbol{\epsilon}.
\end{equation}
Its variation can be cast into the following compact form
\begin{equation}
    \delta\mathbf L=\eps(E_g^{ab}\delta g_{ab}+E_A^a\delta A_a\bigr)
                 +d\mathbf\Theta.
\end{equation}
Here
\begin{equation}
    E_g^{ab}=-\frac{1}{16\pi}\left[G^{ab}-2(F^a_{\;\;c} F^{bc}-\tfrac14g^{ab}F_{cd}F^{cd})\right], \quad
     E_A^a=\frac{\boldsymbol{1}}{4\pi}\nabla_bF^{ba}
\end{equation}
with $E_g^{ab}=0$ and $E_A^a=0$ the Einstein equation and Maxwell equation, respectively. In addition, the bulk symplectic potential is given by
\begin{equation}
    \mathbf{\Theta}=\frac{1}{16\pi}\bigl(g^{ac}\nabla^d\delta g_{cd}-g^{bc}\nabla^a\delta g_{bc}-4F^{ab}\delta A_b\bigr)\boldsymbol{\epsilon}_{a\cdot\cdot\cdot}.
\end{equation}
There are two kinds of local symmetries associated with our Einstein-Maxwell system. 
One arises from the $U(1)$ gauge transformation generated by $\lambda$, which induces a vector field in the covariant phase space as $X_\lambda=\int d^4x\sqrt{-g}(d\lambda)_a\frac{\delta}{\delta A_a}$. The other comes from the diffeomorphism generated by $\xi^a$ and the induced  vector field in the covariant phase space is given by 
$X_\xi=\int d^4x\sqrt{-g}(\mathcal{L}_\xi g_{ab}\frac{\delta}{\delta g_{ab}}+\mathcal{L}_\xi A_a\frac{\delta}{\delta A_a})$.
Then the corresponding Noether currents can be defined as 
\begin{equation}\label{def}
 \mathbf{J}_\lambda\equiv X_\lambda\cdot \mathbf{\Theta},\quad  \mathbf J_\xi\equiv
X_\xi\cdot\mathbf\Theta-\xi\cdot\mathbf L,
\end{equation}
respectively. It turns out that they are closed on shell. In particular, we have 
\begin{equation}\label{closed}
    \mathbf{J}_\lambda=d\mathbf{Q}_\lambda,\quad \mathbf{J}_\xi=d\mathbf{Q}_\xi
\end{equation}
with 
\begin{equation}
\mathbf{Q}_\lambda=-\frac{\lambda}{4\pi}\star\mathbf{F},\quad
 \mathbf Q_\xi=-\frac{1}{16\pi}\star d\boldsymbol{\xi}
 -\frac{\xi\cdot\mathbf A}{4\pi}\star\mathbf F,\label{eq:charge}
\end{equation}
the corresponding Noether charges, where $\star$ denotes the Hodge dual. 

Let $\Gamma$ be a fixed timelike boundary with its outward unit normal $n^a$,
the induced metric $h_{ab}=g_{ab}-n_an_b$, and the induced volume
$\heps=n\cdot\eps$, then we can decompose the symplectic potential at $\Gamma$ as follows
\begin{equation}\label{dp}
    \mathbf\Theta|_\Gamma=-\delta\boldsymbol{\mathcal B}+d\boldsymbol{\mathcal C}+\boldsymbol{\mathcal F}
\end{equation}
with
\begin{equation}
    \boldsymbol{\mathcal B}=\frac{\mathcal K}{8\pi}\heps,\qquad
     \boldsymbol{\mathcal C}=-\frac{1}{16\pi}h^{ac}n^d\delta g_{cd}\heps_{a\cdot\cdot\cdot},\qquad 
     \boldsymbol{\mathcal F}=-\left(\frac{\mathcal K^{ab}-\mathcal K h^{ab}}{16\pi}\delta h_{ab}
 +\frac{\mathcal E^a}{4\pi}\delta A_a\right)\heps,
\end{equation}
where 
$\mathcal K_{ab}=h_a{}^ch_b{}^d\nabla_cn_d$ is the extrinsic curvature and $\mathcal E^a=n_bF^{ba}$. Now suppose that $\Sigma$ is a spacelike hypersurface that emanates from the inner boundary $\partial\Sigma_0$ if it exists and stretches out all the way to $\Gamma$ with the intersection $\Sigma\cap \Gamma$ as its outer boundary $\partial\Sigma_1$. In addition, the induced volume on $\Sigma$ is chosen to be $\bar{\eps}=\nu\cdot\eps$ with $\nu^a$ the future directed unit normal to $\Sigma$ while the induced volumes on its boundaries $\partial\Sigma_0$ and $\partial\Sigma_1$ are further given by the contraction of the outward normal tangential to $\Sigma$ with $\bar{\eps}$. Then we can define the symplectic form as
\begin{equation}
    \Omega\equiv\int_\Sigma\delta\mathbf\Theta-\int_{\partial\Sigma_1}\delta\boldsymbol{\mathcal C}.
\end{equation}
Accordingly, it is not hard to obtain the following fundamental identities
\begin{equation}
   -X_\lambda\cdot\Omega=-\int_{\partial\Sigma_0}\delta\mathbf{Q}_\lambda+\int_{\partial\Sigma_1}\delta\mathbf{Q}_\lambda,\quad  -X_\xi\cdot\Omega=-\int_{\partial\Sigma_0}\bigl(\delta \mathbf{Q}_\xi-\xi\cdot\mathbf{\Theta}\bigr)+\delta H_\xi-\int_{\partial\Sigma_1}\xi\cdot\boldsymbol{\mathcal F}\label{funda}
\end{equation}
for the on shell variations, where $\lambda$ and $\xi$ tangential to $\Gamma$ are both fixed with
\begin{equation}
 H_\xi=\int_{\partial\Sigma_1}\bigl(\mathbf Q_\xi+\xi\cdot\boldsymbol{\mathcal B}
                  -X_\xi\cdot\boldsymbol{\mathcal C}\bigr)\label{eq:Hdef}
\end{equation}
the Hamiltonian conjugate to $\xi$. Note that we have 
\begin{equation}\label{one}
    \int_\Gamma \mathbf{J}_\xi+X_\xi\cdot\delta\boldsymbol{\mathcal{B}}=\int_\Gamma d(X_\xi\cdot \boldsymbol{\mathcal{C}})+X_\xi\cdot \boldsymbol{\mathcal{F}},
\end{equation}
where we have used the fact that $\xi$ is tangential to $\Gamma$.
On the other hand, we also have 
\begin{equation}\label{another}
    \int_\Gamma \mathbf{J}_\xi+X_\xi\cdot\delta\boldsymbol{\mathcal{B}}=d(\mathbf{Q}_\xi+\xi\cdot\boldsymbol{\mathcal{B}}),
\end{equation}
where we have used the diffeomorphism covariance of $\boldsymbol{\mathcal{B}}$ in the sense that $\mathcal{L}_{X_\xi}\boldsymbol{\mathcal{B}}=\mathcal{L}_\xi\boldsymbol{\mathcal{B}}$. Combining Eq. (\ref{one}) and Eq. (\ref{another}) and employing the arbitrariness of $\xi$, one can obtain the following identity
\begin{equation}
    H_\xi=\int_{\partial\Sigma_1}q_\xi\cdot\hat{\eps}
\end{equation}
with
\begin{equation}
  q_\xi^a=-\bigg[\frac{(\mathcal{K}^{ab}-\mathcal{K}h^{ab})\xi_b}{8\pi}+\frac{\mathcal{E}^a\xi\cdot A}{4\pi}\bigg].  
\end{equation}


With the above preparation, let us next consider a stationary black hole with the bifurcation surface $\mathcal{H}$ of its Killing horizon serving as the inner boundary $\partial\Sigma_0$. Then by Eq. (\ref{def}) and Eq. (\ref{closed}), we have $d\mathbf{Q}_\lambda=0$ for a constant $\lambda$, implying
\begin{equation}
   \frac{1} {4\pi}\int_\mathcal{H}\star\mathbf{F}=\frac{1}{4\pi}\int_{\partial\Sigma_1}\star\mathbf{F},
\end{equation}
which we denote by $Q_e$, the electric charge carried by the black hole. Eq. (\ref{funda}) further tells us that the variation of such an electric charge also does not depend on whether we evaluate it on $\mathcal{H}$ or $\partial\Sigma_1$. On the other hand,  it follows from Eq. (\ref{def}), Eq. (\ref{closed}), Eq. (\ref{eq:charge}), and Eq. (\ref{eq:Hdef}) that 
\begin{equation}\label{1}
    \beta(H_\xi-TS-\Phi_\mathcal{H}Q_e)=I_E,
\end{equation}
for the Killing vector field $\xi$ normal to the horizon and satisfying $\mathcal{L}_\xi\mathbf{A}=0$,
where the inverse Hawking temperature $\beta=\frac{1}{T}=\frac{2\pi}{\kappa}$ with $\kappa$ the surface gravity of the black hole, the horizon electric potential $\Phi_\mathcal{H}=-(\xi\cdot \mathbf{A})|_{\mathcal H}$, and 
\begin{equation}
   S=-\frac{1}{8\kappa}\int_\mathcal{H}\star d\boldsymbol{\xi}, \quad I_E=\int_M\mathbf{L}_E+\int_\Gamma  \boldsymbol{\mathcal B}_E
\end{equation}
denote the Wald entropy and Euclidean on shell action of the black hole, respectively.  
In addition, Eq. \eqref{funda} gives rise to 
\begin{equation}
    T\delta S+\Phi_\mathcal{H}\delta Q_e=\delta H_\xi-\int_{\partial\Sigma_1}\xi\cdot\boldsymbol{\mathcal F}.\label{fun2}
\end{equation}
Now comes the background subtraction method. As such, we are required to introduce a reference spacetime without an inner boundary and carrying no electric charge. For such a reference spacetime, the counterparts of Eq. (\ref{1}) and Eq. (\ref{fun2}) read 
\begin{equation}
    \beta H_\xi^0=I_E^0,\quad 0=\delta H^0_\xi-\int_{\partial\Sigma_1}\xi\cdot\boldsymbol{\mathcal F}^0
\end{equation} 
where the superscript $0$ denotes the corresponding quantities evaluated on the reference spacetime. Using the square bracket to represent the difference from the reference spacetime. For instance, $[H_\xi]\equiv H_\xi-H_\xi^0$ and $[\boldsymbol{\mathcal F}]\equiv\boldsymbol{\mathcal F}-\boldsymbol{\mathcal F}^0$. Then the criterion for the applicability of the background subtraction method is given by 
\begin{equation}
    \int_{\partial\Sigma_1}\xi\cdot[\boldsymbol{\mathcal F}]=0,\label{criterion}
\end{equation}
and the finiteness of $[H_\xi]$.
With this criterion satisfied, not only do we have the first law of black hole thermodynamics
\begin{equation}
    \frac{\kappa}{2\pi}\delta S+\Phi_\mathcal{H}\delta Q_e+\Omega_\mathcal{H}\delta J=\delta M,
\end{equation}
but also we have the so called quantum statistical relation
\begin{equation}
    \beta G\equiv \beta(M-\Omega_\mathcal{H}J-TS-\Phi_\mathcal{H}Q_e)=[I_E]
\end{equation}
with $G$ the Gibbs free energy, 
where we have used the standard decomposition $\xi=\frac{\partial}{\partial t}+\Omega_\mathcal{H}\frac{\partial}{\partial\varphi}$ as well as the definition for the mass and angular momentum, namely $M=[H_\frac{\partial}{\partial t}]$ and $J=-[H_{\frac{\partial}{\partial\varphi}}]$.

We end this section by some remarks. First, Eq. (\ref{criterion}) amounts to the integrability condition by requiring the closedness of $\int_{\partial\Sigma_1}\xi\cdot\boldsymbol{\mathcal{F}}$ because of $\int_{\partial\Sigma_1}\xi\cdot\boldsymbol{\mathcal{F}}=\delta H_\xi^0$. In addition, although $\partial\Sigma_1$ is usually chosen to approach the spatial infinity in the background subtraction method for the conventional asymptotics, the check of such an integrability condition does not depend on where we locate $\partial \Sigma_1$ due to Eq. (\ref{fun2}). Last but not least, the integrability condition presented in \cite{Barnnich2002,Barnnich2003,Barnnich2008} amounts to requiring the closedness of $\int_{\partial\Sigma_1}\xi\cdot\mathbf{\Theta}$, which reduces to our form of integrability condition when restricted onto the stationary variations according to the decomposition made in Eq. (\ref{dp})\footnote{For the stationary variations, we have $\int_{\partial\Sigma_1}\xi\cdot\mathbf{\Theta}=\delta (\int_{\partial\Sigma_1}\xi\cdot\boldsymbol{\mathcal{B}})+\int_{\partial\Sigma_1}\xi\cdot\boldsymbol{\mathcal{F}}$, because $\xi\cdot d\boldsymbol{\mathcal{C}}=\mathcal{L}_\xi\boldsymbol{\mathcal{C}}-d(\xi\cdot\boldsymbol{\mathcal{C}})=-d(\xi\cdot\boldsymbol{\mathcal{C}})$ does not contribute to the integral over $\partial\Sigma_1$.}. 
It is due to such a decomposition that our form of integrability condition turns out to be much easier for one to check.


\section{The Kerr-Newman-Bertotti-Robinson black holes}\label{sec:KN-BR}

The KNBR black holes, as a family of solutions to the Einstein-Maxwell equations with the four free parameters $(m,a,e,B)$, take the following form \cite{KNBR}
\begin{eqnarray}
    &&ds^2=\frac{1}{\Omega^2}\bigg[
 -\frac{Q}{\rho^2}\left(d\tilde t-
          \frac{a\sin^2\theta}{K}d\tilde\varphi\right)^2
 +\frac{\rho^2}{Q}d r^2+\frac{\rho^2}{P}d\theta^2
 +\frac{P\sin^2\theta}{\rho^2}\left(ad\tilde t-
          \frac{r^2+a^2}{K}d\tilde\varphi\right)^2\bigg],\nonumber\\
          &&{\mathbf A}=\frac{a}{\Omega}\left[-(1+k^2)B+
            \frac{r(mB\cos^2\theta-k)}{\rho^2}\right]d\tilde t\nonumber\\
            &&\qquad+\frac{1}{K\Omega}\left[(1+k^2)a^2B-kr-\frac1B
 +\frac{r}{\rho^2}\left(-mB(r^2+a^2)\cos^2\theta
    +ka^2\sin^2\theta\right)\right]d\tilde\varphi+d\lambda,\label{potential}
\end{eqnarray}
where
\begin{align}
 \rho^2&=r^2+a^2\cos^2\theta,&P&=1+B^2\mu^2\cos^2\theta,
 &I&=(1+kBr)^2+B^2r^2,\nonumber\\
 \Delta&=(1+k^2)a^2-2mr+(1+k^2-\frac{e^2}{a^2}) r^2,&Q&=I\Delta,
 &\Omega^2&=I-B^2\Delta\cos^2\theta\nonumber
\end{align}
with
\begin{align}
     K&=1+B^2\mu^2,\qquad\;
     &\mu&=\sqrt{m^2-(1+k^2)^2a^2+(1+k^2)e^2},\qquad
     \nonumber\\
     k&=\frac{es-amB}{a(1+e^2B^2)},\qquad
     &s&=\sqrt{1+(e^2-a^2)B^2-(m^2+e^2)a^2B^4}.
\end{align}
$d\lambda$ in Eq. \eqref{potential} represents the gauge transformation. For our purpose, we restrict ourselves onto the gauge freedom in which $\lambda$ is the function of time and azimuthal angle and $d\lambda$ is invariant under time translation and rotation\footnote{The gauge freedom with $\lambda$ as an arbitrary function of $(r,\theta)$ is irrelevant in the following discussions.}. As a result, 
\begin{equation}
    d\lambda=\alpha d\tilde t+\chi d\tilde\varphi
\end{equation}
up to the undetermined constants $\alpha$ and $\chi$. To make our life easier, $\chi$ can be further fixed as follows \cite{ZZC}
\begin{equation}
    \chi=\frac{s+ameB^3}{(1+e^2B^2)BK}
    \end{equation}
by requiring the resulting $\mathbf A$ regular on the rotating axis, namely $A_{\tilde\varphi}|_{\theta=0,\pi}=0$. Thus we are left with one gauge freedom characterized by $\alpha$. On the other hand, we also have the freedom to choose the time generator 
\begin{equation}
    \frac{\partial}{\partial t}=\gamma(\frac{\partial}{\partial\tilde{t}}+\omega\frac{\partial}{\partial \tilde{\varphi}})
\end{equation}
with $\gamma$ and $\omega$ undetermined constants, which can be implemented by the following coordinate transformation
\begin{equation}
 t=\frac{1}{\gamma}\tilde t,\qquad
 \varphi=\tilde\varphi-\omega \tilde t.
\end{equation}
Accordingly, the KNBR black holes read
\begin{align}
 ds^2=&\frac{1}{\Omega^2}\bigg\{
 -\frac{Q}{\rho^2}
 \left[
 \gamma\left(1-\frac{a\omega\sin^2\theta}{K}\right)dt
 -\frac{a\sin^2\theta}{K}d\varphi
 \right]^2
+\frac{P\sin^2\theta}{\rho^2}
 \left[
 \gamma\left(a-\frac{(r^2+a^2)\omega}{K}\right)dt
 -\frac{r^2+a^2}{K}d\varphi
 \right]^2\nonumber\\
 &\qquad\;\;+\frac{\rho^2}{Q}dr^2
 +\frac{\rho^2}{P}d\theta^2
 \bigg\},
\nonumber\\
 {\mathbf A}=&\bigg\{\frac{a\gamma}{\Omega}\left[-(1+k^2)B+
            \frac{r(mB\cos^2\theta-k)}{\rho^2}\right]+\gamma\alpha \bigg\}dt\nonumber\\
            &+\bigg\{\frac{1}{K\Omega}\left[(1+k^2)a^2B-kr-\frac1B
 +\frac{r}{\rho^2}\left(-mB(r^2+a^2)\cos^2\theta
    +ka^2\sin^2\theta\right)\right]+\frac{s+ameB^3}{(1+e^2B^2)BK}\bigg\}(d\varphi+\gamma\omega dt).
 \label{eq:adaptedpotential}
\end{align}

Without loss of generality but for simplicity, below we shall focus on the KNBR black hole with $B\ge 0$. 
The corresponding event horizon is located at
\begin{equation}
 r_\pm=\frac{m\pm\mu}{1+k^2-e^2/a^2},
 \label{eq:outerhorizon}
\end{equation}
whereby we further have
\begin{eqnarray}
    &&\Omega_\mathcal{H}=-\frac{g_{t\varphi}}{g_{\varphi\varphi}}\bigg|_{r_+}=\gamma\left(\frac{aK}{r_+^2+a^2}-\omega\right),\quad \Phi_\mathcal{H}=\gamma\left(\frac{er_+}{r_+^2+a^2}-\alpha\right), \quad T
     =\frac{\gamma\mu I(r_+)}{2\pi(r_+^2+a^2)},\quad 
 S=\frac{\pi(r_+^2+a^2)}{K I(r_+)}.\label{eq:horizonquantities}
\end{eqnarray}
In addition, the electric charge carried by the black hole is given by 
\begin{equation}
    Q_e=\frac{e}{K}.
\end{equation}
For $B=0$, corresponding to the asymptotically flat KN black holes, $\alpha$ is usually chosen to be zero by requiring the electric potential vanishes at the spatial infinity, while $\gamma$ and $\omega$ are fixed respectively to be one and zero by choosing the non-rotating observers at the spatial infinity on physical grounds. However, for $B\neq 0$, the corresponding black holes are not asymptotically flat, leading to the ambiguity in gauging the electric potential and choosing the time generator. But nevertheless, as we shall show in the next section, such an ambiguity can be resolved by the integrability condition together with the above gauge choice for the KN black holes as the initial condition.

\section{Background subtraction and black hole thermodynamics}
\label{sec:subtraction}
To start off, we choose $\Sigma$ to be a constant $t$ surface and $\Gamma$ to be $r=R$. As remarked at the end of Section \ref{sec:formalism}, the criterion for the applicability of the background subtraction method implies the following integrability condition
\begin{equation}
\delta(\int_{\partial\Sigma_1}\xi\cdot\boldsymbol{\mathcal{F}})=0,
\end{equation}
on the moduli space parameterized by $(m,e,a,B)$. As explicitly calculated in Appendix \ref{app:boundary} and \ref{app:moduli-integrability}, $\int_{\partial\Sigma_1}\xi\cdot\boldsymbol{\mathcal{F}}$ can be cast into the following compact form
\begin{equation}
     \int_{\partial\Sigma_1}\xi\cdot\boldsymbol{\mathcal F}
 =\delta H_{\frac{\partial}{\partial t}}-\mathcal A,
\end{equation}
where $\mathcal{A}$ is an $R$ independent one-form on the moduli space and can be written as follows
\begin{equation}
 \mathcal{A}=\gamma\nu^1+\gamma\omega\nu^2+\gamma\alpha\nu^3
\end{equation}
with $\nu^i$ the three linearly independent one-forms. Accordingly, our integrability condition reduces to 
\begin{equation}
    \delta\mathcal{A}=0,
\end{equation}
which is $R$ independent. As demonstrated in Appendix \ref{app:moduli-integrability}, this equation can be massaged as the three decoupled first-order ordinary differential equations for $\alpha$, $\gamma$, and $\omega$ along the integral curves of the characteristic vector field 
    \begin{equation}
T^A=\epsilon^{ABCD} (\nu^1\wedge\nu^2\wedge\nu^3)_{BCD}
\end{equation}
on the moduli space with $\epsilon^{ABCD}$ the appropriately normalized contravariant volume. By viewing $B$ as the time function and taking 
\begin{equation}
    \alpha|_{B=0}=0,\quad \gamma|_{B=0}=1,\quad \omega|_{B=0}=0
\end{equation}
for the KN black holes as the initial data, the three equations can be solved uniquely as follows
\begin{align} \alpha&=
 -\frac{e^2\left(\frac ea-k\chi BK\right)}{a(m(\chi BK)^2+e^2kB)},\nonumber\\
    \gamma&=\frac{m(\chi BK)^2+e^2kB}{\sqrt{K(
 m^2+e^2K-a^2k^2(1+k^2-\frac{e^2}{a^2}))
 }},
\nonumber\\
 \omega&=
 -\frac{
 aKB\left[
 k\left(1+k^2-\frac{e^2}{a^2}\right)
 +(1+k^2)mB
 \right]
 }{m(\chi BK)^2+e^2kB}.
 \label{goa}
\end{align}
Accordingly, it is not hard to show $\mathcal A=\delta f$ with
\begin{equation}
f=
\frac{\sqrt{m^2+e^2(K+k^2)-a^2k^2(1+k^2)}}{K^{3/2}}.
\end{equation}

Among others, below we shall further show that the background subtraction method enables one to obtain the expected mass and angular momentum for the KNBR black hole.
As such, we take the Minkowski spacetime
\begin{equation}
ds^2_0=-\eta_R^2 dt^2
+\frac{r^2+a^2\cos^2\theta}{r^2+a^2}\,dr^2
+(r^2+a^2\cos^2\theta)\,d\theta^2+(r^2+a^2)\sin^2\theta\,d\varphi^2.
\label{bgmetric}
\end{equation}
as the reference spacetime, where $\eta_R$ is an $R$ dependent time scaling factor, to be determined later on. Note that now we have 
\begin{equation}
    \delta H_\xi^0=
    \int_{\partial\Sigma_1} \xi\cdot\boldsymbol{\mathcal F}^0=
    \int_{\partial\Sigma_1} \xi\cdot\boldsymbol{\mathcal F}=
    \delta H_{\frac{\partial}{\partial t}}-\mathcal A=\delta(H_{\frac{\partial}{\partial t}}-f),
\end{equation}
which implies that 
\begin{equation}
    H_\xi^0=H_{\frac{\partial}{\partial t}}-f+C
    \label{criterion2}
\end{equation}
with $C$ an undetermined constant on the moduli space. 
On the other hand, we also have
\begin{equation}
H^0_{\frac{\partial}{\partial t}}
=-\frac{\eta_R}{2}
\left[
R+\frac{R^2+a^2}{a}\arctan\!\left(\frac aR\right)
\right],
\quad H^0_{\frac{\partial}{\partial\varphi}}=0
\end{equation}
by using Eq. (\ref{appendixformula}) with $m=e=B=0$ and $\gamma=\eta_R$. Thus we have the mass 
\begin{equation}
    M=[H_{\frac{\partial}{\partial t}}]=f-C=f,
\end{equation}
where $C=0$ follows from the requirement that $M=m$ for the KN black hole. In addition, the scaling factor can also be fixed as

\begin{equation}
    \eta_R=\frac{2\big(M-H_{\frac{\partial}{\partial t}}\big)}{R+\frac{R^2+a^2}{a}\arctan(\frac aR)}.
\end{equation}
The angular momentum is given simply by 
\begin{equation}
    J=-[H_\frac{\partial}{\partial \varphi}]=\frac{ma}{K^2},
\end{equation}
where Eq.(\ref{appendixformula}) is used again. It is noteworthy that our angular momentum recovers the result for the KN black hole.

As stated in Section \ref{sec:formalism}, not only the first law of black hole thermodynamics but also the quantum statistical relation holds automatically for the thermodynamical quantities we have obtained. In addition, one can check that they also obey the CR relation 
\begin{equation}
 M^2=\frac{S}{4\pi}+\frac{Q_e^2}{2}
                  +\frac{\pi(Q_e^4+4J^2)}{4S},
 \label{eq:massrelation}
\end{equation}
and the Smarr formula
\begin{equation}
 M=2TS+2\Omega_\mathcal{H}J+\Phi_\mathcal{H}Q_e.
 \label{eq:thermodynamiclaws}
\end{equation}

\section{Conclusion}
The recently found KNBR black holes are not asymptotically flat when the magnetic field is present.
This leads to the ambiguity in the gauge choice for the time generator and the electric potential. By using the newly established form of integrability condition together with the gauge choice for the KN black holes as the initial condition, we have removed such an ambiguity. We further obtain the mass and angular momentum by the background subtraction method with the Minkowski spacetime as the reference spacetime. As a result, not only is the first law of black hole thermodynamics as well as the quantum statistical relation automatically satisfied, but also both the CR relation and the Smarr formula are obeyed.

We conclude our paper with some discussions. For one thing, we have succeeded in resolving the deficiencies in the previous exploration of the thermodynamic properties for the KNBR black hole, where some ad hoc assumptions were made. For another, the background subtraction method is usually used to remove the divergence arising in the spatial infinity with conventional asymptotics. Our work demonstrates that the background subtraction method also displays its remarkable power in addressing the thermodynamic issues related to the black holes with unconventional asymptotics. In particular, the involved cut-off surface is not required to be approaching the spatial infinity. Instead, it can be located in the vicinity of the black hole, and even allowed to approach the black hole horizon, echoing the treatment for the isolated horizon. On the other hand, we have been adopting the form of the integrability condition established in \cite{RefA}, which can be regarded as the improved version on that presented in \cite{Barnnich2002,Barnnich2003,Barnnich2008}. In addition,  we obtain the mass by first fixing the gauge in a straightforward and transparent manner rather than resorting to some well educated guess as made in \cite{Barnnich2005,Barnnich2006,Compere2007,Astorino2016,Li2026}.
So it is worthwhile to apply our framework and strategy to investigate the thermodynamic properties of the other black holes with unconventional asymptotics, including those found recently in \cite{Muller}. We hope to address these issues elsewhere in the future. 

\section*{Acknowledgements}

This work is partially supported by the National Natural Science Foundation of China with Grant Nos. 12361141825,  12375058, and 12575047. H.Z. likes to acknowledge Zhengwen Liu for his fantastic hospitality at Shing-Tung Yau Center of Southeast University during the final stage of this work.

\appendix
\section{Explicit computation of \(H_\xi\) and \texorpdfstring{\(\int_{\partial\Sigma_1}\xi\cdot \boldsymbol{\mathcal{F}}\)}{} at $\Gamma$}
\label{app:boundary}
In order to evaluate $H_\xi$ and $\int_{\partial\Sigma_1}\xi\cdot\boldsymbol{\mathcal{F}}$ at $\Gamma$ for $\xi=\frac{\partial}{\partial t}+\Omega_\mathcal{H}\frac{\partial}{\partial\varphi}$, we like to express the metric of KNBR black holes as
$ds^2=g_{rr}dr^2+h_{ij}dx^i dx^j$ with $i,j=t,\theta,\varphi$. As a result, we have
\begin{align}
 &H_{\frac{\partial}{\partial t}}
 =\int_0^\pi d\theta\,\sqrt{|h|}
 \left[
 \frac{\mathcal K^t{}_t-\mathcal K}{4}
 +\frac{A_t\mathcal E^t}{2}
 \right]\Bigg |_R,\qquad
 H_{\frac{\partial}{\partial\varphi}}
 =\int_0^\pi d\theta\,\sqrt{|h|}
 \left[
 \frac{\mathcal K^t{}_\varphi}{4}
 +\frac{A_\varphi\mathcal E^t}{2}
 \right]\Bigg|_R,
 \label{eq:app-Hdefinition}\\
 &\int_{\partial\Sigma_1}\xi\cdot\boldsymbol{\mathcal F}
 =\int_{\partial\Sigma_1}\frac{\partial}{\partial t}\cdot\boldsymbol{\mathcal F}
 =\int_0^\pi d\theta\,\sqrt{|h|}
 \left[
 \frac{\mathcal K^{ij}-\mathcal K h^{ij}}{8}\delta h_{ij}
 +\frac{\mathcal E^i}{2}\delta A_i
 \right]\Bigg|_R.
 \label{eq:app-Idefinition}
\end{align}
The non-vanishing components of the induced metric and its inverse read
\begin{align}
 h_{tt}
 ={}&
 \frac{\gamma^2}{\Omega^2\rho^2}
 \left[
 -Q\left(1-\frac{a\omega\sin^2\theta}{K}\right)^2
 +P\sin^2\theta
 \left(a-\frac{\omega(R^2+a^2)}{K}\right)^2
 \right],
\nonumber
 \\
 h_{t\varphi}
 ={}&
 \frac{\gamma\sin^2\theta}{K\Omega^2\rho^2}
 \left[
 aQ\left(1-\frac{a\omega\sin^2\theta}{K}\right)
 -P(R^2+a^2)
 \left(a-\frac{\omega(R^2+a^2)}{K}\right)
 \right],\nonumber
 \\
 h_{\varphi\varphi}
 ={}&
 \frac{\sin^2\theta}{K^2\Omega^2\rho^2}
 \left[P(R^2+a^2)^2-a^2Q\sin^2\theta\right],
 \qquad\qquad
 h_{\theta\theta}=\frac{\rho^2}{P\Omega^2},\nonumber\\
h^{tt}
={}&
\frac{\Omega^2}{\gamma^2\rho^2}
\left[
-\frac{(R^2+a^2)^2}{Q}
+\frac{a^2\sin^2\theta}{P}
\right],\nonumber
\\
h^{t\varphi}
={}&
\frac{\Omega^2}{\gamma\rho^2}
\left[
-\frac{(R^2+a^2)\bigl[aK-\omega(R^2+a^2)\bigr]}{Q}
+\frac{a\bigl(K-a\omega\sin^2\theta\bigr)}{P}
\right],\nonumber
\\
h^{\varphi\varphi}
={}&
\frac{\Omega^2}{\rho^2}
\left[
-\frac{\bigl[aK-\omega(R^2+a^2)\bigr]^2}{Q}
+\frac{\bigl(K-a\omega\sin^2\theta\bigr)^2}
{P\sin^2\theta}
\right],
\qquad
h^{\theta\theta}
={}\frac{P\Omega^2}{\rho^2},
 \label{eq:app-KNBR-finite-h}
\end{align}
with $\sqrt{|h|}
=\frac{\gamma\rho\sqrt Q\,\sin\theta}{K\Omega^3}$.
The nonvanishing components of the  extrinsic curvature  are given by
\begin{align}
 \mathcal K_{tt}
 ={}&
 \frac{\gamma^2\sqrt Q}{2\Omega\rho^3}
 \Bigg\{
 -\partial_rQ
 \left(1-\frac{a\omega\sin^2\theta}{K}\right)^2
 -\frac{4R\omega P\sin^2\theta}{K}
 \left(a-\frac{\omega(R^2+a^2)}{K}\right)
 \nonumber\\
 &\qquad\qquad\;\:
 -\left[
 -Q\left(1-\frac{a\omega\sin^2\theta}{K}\right)^2
 +P\sin^2\theta
 \left(a-\frac{\omega(R^2+a^2)}{K}\right)^2
 \right]
 \left[
 \frac{2R}{\rho^2}
 +\frac{\partial_r(\Omega^2)}{\Omega^2}
 \right]
 \Bigg\},\nonumber
 \\
 \mathcal K_{t\varphi}
 ={}&
 \frac{\gamma\sin^2\theta\sqrt Q}{2K\Omega\rho^3}
 \Bigg\{
 a\,\partial_rQ
 \left(1-\frac{a\omega\sin^2\theta}{K}\right)
 -2RP\left(a-\frac{2\omega(R^2+a^2)}{K}\right)
 \nonumber\\
 &\qquad\qquad\qquad
 -\left[
 aQ\left(1-\frac{a\omega\sin^2\theta}{K}\right)
 -P(R^2+a^2)
 \left(a-\frac{\omega(R^2+a^2)}{K}\right)
 \right]
 \left[
 \frac{2R}{\rho^2}
 +\frac{\partial_r(\Omega^2)}{\Omega^2}
 \right]
 \Bigg\},\nonumber
 \\
 \mathcal K_{\varphi\varphi}
 ={}&
 \frac{\sin^2\theta\sqrt Q}{2K^2\Omega\rho^3}
 \Bigg\{
 4RP(R^2+a^2)-a^2\sin^2\theta\,\partial_rQ
 -\left[P(R^2+a^2)^2-a^2Q\sin^2\theta\right]
 \left[
 \frac{2R}{\rho^2}
 +\frac{\partial_r(\Omega^2)}{\Omega^2}
 \right]
 \Bigg\},\nonumber
 \\
 \mathcal K_{\theta\theta}
 ={}&
 \frac{\sqrt Q}{2P\Omega\rho}
 \left[
 2R-\rho^2\frac{\partial_r(\Omega^2)}{\Omega^2}
 \right],
 \label{eq:app-KNBR-finite-extrinsic}
\end{align}
where
\begin{align}
 \partial_rQ
 ={}&
 2B\left[k+(1+k^2)BR\right]\Delta
 +2I\left[
 \left(1+k^2-\frac{e^2}{a^2}\right)R-m
 \right],\nonumber
 \\
 \partial_r(\Omega^2)
 ={}&
 2B\left[k+(1+k^2)BR\right]
 -2B^2\left[
 \left(1+k^2-\frac{e^2}{a^2}\right)R-m
 \right]\cos^2\theta .
 \label{eq:app-KNBR-radial-derivatives}
\end{align}
In addition, we also have the non-vanishing components of electromagnetic potential and $\mathcal E$ as follows
\begin{align}
 A_\varphi
 ={}&
 \chi+\frac1{K\Omega}
 \left[
 (1+k^2)a^2B-kR-\frac1B
 +\frac R{\rho^2}
 \left(
 -mB(R^2+a^2)\cos^2\theta
 +ka^2\sin^2\theta
 \right)
 \right],\nonumber
 \\
 A_t
 ={}&
 \gamma\Bigg\{
 \alpha
 +\frac a\Omega
 \left[
 -(1+k^2)B
 +\frac{R(mB\cos^2\theta-k)}{\rho^2}
 \right]
 +\omega\chi
 \nonumber\\
 &\qquad
 +\frac{\omega}{K\Omega}
 \left[
 (1+k^2)a^2B-kR-\frac1B
 +\frac R{\rho^2}
 \left(
 -mB(R^2+a^2)\cos^2\theta
 +ka^2\sin^2\theta
 \right)
 \right]
 \Bigg\},
 \label{eq:app-KNBR-finite-potential}
\end{align}
and
\begin{align}
\mathcal  E^t
 ={}&
 \frac{a(R^2+a^2)}{\gamma\rho\sqrt Q}
 \Bigg\{
 \frac{\Omega^2
 (mB\cos^2\theta-k)(R^2-a^2\cos^2\theta)}
 {\rho^4}+\frac{RB\sin^2\theta}{(R^2+a^2)\rho^2}
 \left[Q-P(R^2+a^2)\right]
 \nonumber\\
 &\qquad
 +
 \left[
 -(1+k^2)B
 +\frac{R(mB\cos^2\theta-k)}{\rho^2}
 \right]
 \left[
 B\bigl(k+(1+k^2)BR\bigr)
 -B^2\left(
 \left(1+k^2-\frac{e^2}{a^2}\right)R-m
 \right)\cos^2\theta
 \right]
 \Bigg\},\nonumber
 \\
 \mathcal E^\varphi
 ={}&
 \gamma\left(\frac{aK}{R^2+a^2}-\omega\right)\mathcal E^t
 +\frac{KBR\sqrt Q}{\rho(R^2+a^2)} .
 \label{eq:app-KNBR-finite-electric}
\end{align}

The integral in \eqref{eq:app-Hdefinition} and \eqref{eq:app-Idefinition} involves the following form
\begin{equation}
 \mathcal I_n(D,E)
 \equiv
 \int_0^\pi
 \frac{\sin\theta\,d\theta}
 {(D-E\cos^2\theta)^n}
 =
 \int_{-1}^{1}\frac{dx}{(D-Ex^2)^n},
 \qquad D>0,\quad D-E>0.
 \label{eq:app-J-family}
\end{equation}
with $n$ being a positive integer or a half-integer. 
$\mathcal{I}_n$ can be calculated out through the following recursion relation
\begin{equation}
 \mathcal I_n(D,E)
 =
 \frac{1}{(n-1)D(D-E)^{n-1}}
 +
 \frac{2n-3}{2(n-1)D}\,
 \mathcal I_{n-1}(D,E),
 \qquad n>1
 \label{eq:app-J-recurrence}
\end{equation}
with
\begin{align}
 \mathcal I_1(D,E)
 &=
 \begin{cases}
 \displaystyle
 \frac{2}{\sqrt{DE}}
 \operatorname{artanh}\sqrt{\frac ED},
 & E>0,\\[2mm]
 \displaystyle \frac2D,
 & E=0,\\[2mm]
 \displaystyle
 \frac{2}{\sqrt{-DE}}
 \arctan\sqrt{\frac{-E}{D}},
 & E<0,
 \end{cases}\nonumber
 \\
 \mathcal I_{3/2}(D,E)
 &=\frac{2}{D\sqrt{D-E}}.
 \label{eq:app-J-base-values}
\end{align}
Then a straightforward calculation gives us the following compact result 
\begin{align}\label{appendixformula}
H_{\frac{\partial}{\partial t}}
={}&\frac{\gamma}{2K}\Bigg\{
m-\left(1+k^2-\frac{e^2}{a^2}\right)R +\frac{
a^2\left\{mBR^2I-k\left[a^2I+2B^2R^2\Delta\right]\right\}}
{B(R^2+a^2)(a^2I+B^2R^2\Delta)}
+\frac{
e\left[a^2I+B^2(R^2+a^2)\Delta\right]}
{B(R^2+a^2)(a\chi BK+eBR)}\nonumber
\\
&+\frac{BR^2\Delta}
{\sqrt Q\,(a^2I+B^2R^2\Delta)^2}
\Bigg[
 B\Delta\left[k+(1+k^2)BR\right]
       (3a^2I+B^2R^2\Delta)\nonumber
\\
&\hspace{34mm}
+I\left[\left(1+k^2-\frac{e^2}{a^2}\right)R-m\right]
       (B^2R^2\Delta-a^2I)
-2B^2RI\Delta^2
\Bigg]
\operatorname{artanh}\!\left(B\sqrt{\frac{\Delta}{I}}\right)\nonumber
\\
&-\frac{
aQ\left[a^2I+B^2R^2\Delta-2BR(a^2k+mBR^2)\right]}
{(a^2I+B^2R^2\Delta)^2}
\arctan\!\left(\frac aR\right)\Bigg\}
-\frac{\gamma\omega ma}{K^2}
-\frac{\gamma\alpha e}{K},\nonumber\\
    H_{\frac{\partial}{\partial\varphi}}&=-\frac{ma}{K^2},\nonumber\\
    \int_{\partial\Sigma_1}
 \xi\cdot\boldsymbol{\mathcal F}
 &=\delta H_{\frac{\partial}{\partial t}}-\mathcal A,
\end{align}
where $H_{\frac{\partial}{\partial t}}$ is $R$ dependent while both $H_{\frac{\partial}{\partial \varphi}}$ and $\mathcal{A}$ are $R$ independent with 
\begin{align}
\mathcal A
={}&
\gamma\Bigg[
\frac{\delta m}{K}
+\frac{1-(\chi BK)^2}{BK(\chi BK)}
\,\delta\!\left(\frac ea\right)
-\frac{k}{BK(\chi BK)}\,\delta(\chi BK)
-\frac{1}{2BK^2}
\left(\frac{e}{a(\chi BK)}+2mB-k\right)\delta K
\Bigg]\nonumber
\\
&-\gamma\omega\,\delta\!\left(\frac{ma}{K^2}\right)
-\gamma\alpha\,\delta\!\left(\frac eK\right).
\end{align}


\section{Explicit derivation of $\alpha$, $\gamma$ and $\omega$}
\label{app:moduli-integrability}
$(m, a, e, B)$ serve naturally as the canonical coordinates for the moduli space of the KNBR black holes.  In this appendix, we shall show how  $\alpha$, $\gamma$ and $\omega$ as the functions of the moduli space can be fixed by solving the integrability equation
\begin{equation}
    \delta\bigg(
 \int_{\partial\Sigma_1}
 \xi\cdot\boldsymbol{\mathcal F}\bigg)=0
\end{equation}
subject to the following initial condition 
\begin{equation}
 \alpha|_{B=0}=0,\quad   \gamma|_{B=0}=1,\quad\ \omega|_{B=0}=0
    \label{initial}
\end{equation}
on $B=0$ surface, which corresponds to the standard choice for the KN black holes. 

 As shown in Appendix \ref{app:boundary}, we have
\begin{equation}
 \int_{\partial\Sigma_1}
 \xi\cdot\boldsymbol{\mathcal F}
 =\delta H_{\frac{\partial}{\partial t}}-\mathcal{A}.
 \label{eq:moduli-boundary}
\end{equation}
Here $\mathcal{A}$ can be rewritten as 
\begin{equation}
 \mathcal{A}=\gamma\nu^1+\gamma\omega\nu^2+\gamma\alpha\nu^3,
 \label{eq:moduli-zeta}
\end{equation}
where
\begin{align}
 \nu^1=\nu^1_m\delta m+\nu^1_a\delta a+\nu^1_e\delta e+\nu^1_B\delta B,\quad 
 \nu^2=-\delta\left(\frac{ma}{K^2}\right),\quad
 \nu^3=-\delta\left(\frac{e}{K}\right)
 \label{eq:moduli-nu}
\end{align}
with
\begin{align}
 \nu^1_m={}&
 \frac{
 k\left[1-\left(2(1+k^2)-\frac{e^2}{a^2}\right)a^2B^2\right]
 }{K(e/a)s}
 \nonumber\\
 &+\frac{
 B\left[
 k\left(1+k^2-\frac{e^2}{a^2}\right)
 +\left(1+k^2-\frac{2e^2}{a^2}\right)mB
 \right]
 }{K^2(e/a)^2}
 \left[
 m-\frac{Bk(\chi BK)\left(e^2-2(1+k^2)a^2\right)}{s}
 \right],
 \nonumber\\
 \nu^1_a={}&
 \frac{
 aB\left[
 \left(1+k^2-\frac{e^2}{a^2}\right)(1-kmB)
 -\frac{e^2k^2}{a^2}
 \right]
 }{K(e/a)s}
 \nonumber\\
 &-\frac{
 B\left[
 k\left(1+k^2-\frac{e^2}{a^2}\right)
 +\left(1+k^2-\frac{2e^2}{a^2}\right)mB
 \right]
 }{K^2(e/a)^2}
 \left[
 (1+k^2)^2a
 +\frac{ek\left(e^2-2(1+k^2)a^2\right)}{a^2s}
 \right],
 \nonumber\\
 \nu^1_e={}&
 \frac{
 akB\left[
 \frac{ek}{a}
 +\left(2(1+k^2)-\frac{e^2}{a^2}\right)(\chi BK)
 \right]
 }{K(e/a)s}
 \nonumber\\
 &+\frac{
 B\left[
 k\left(1+k^2-\frac{e^2}{a^2}\right)
 +\left(1+k^2-\frac{2e^2}{a^2}\right)mB
 \right]
 }{K^2(e/a)^2}
 \left[
 (1+k^2)e
 +\frac{k(\chi BK)^2\left(e^2-2(1+k^2)a^2\right)}{as}
 \right],
 \nonumber\\
 \nu^1_B={}&
 \frac{
 -m\left(m+(1+k^2)a^2kB\right)
 +(1+k^2)a^2\left(1+2k^2-\frac{e^2}{a^2}\right)
 }{K(e/a)s}
 \nonumber\\
 &+\frac{
 k\left(1+k^2-\frac{e^2}{a^2}\right)
 +\left(1+k^2-\frac{2e^2}{a^2}\right)mB
 }{K^2(e/a)^2}
 \left[
 \mu^2-\frac{
 Bk\left(e^2-2(1+k^2)a^2\right)
 \left(m\chi BK+(1+k^2)aeB\right)
 }{s}
 \right].
 \label{eq:moduli-bcoefficients}
\end{align}
Since $\delta^2 H_{\frac{\partial}{\partial t}}=0$, the integrability equation reduces to
\begin{equation}
 \delta\mathcal{A}=0
 \label{eq:moduli-closure}
\end{equation}
with the initial condition given by
\begin{equation}\label{iv}
    \mathcal{A}|_{B=0}=\delta m+\frac{e(a^2+e^2)}{a}\delta B.
\end{equation}
Now we construct a characteristic vector field $T$ on the moduli space by 
\begin{equation}
\epsilon^{ABCD} (\nu^1\wedge\nu^2\wedge\nu^3)_{BCD}
\end{equation}
where the normalization of the contravariant volume  is so chosen that the resulting $T$ takes the following form
 \begin{equation}
T=T^m\partial_m+T^a\partial_a+T^e\partial_e+T^B\partial_B,
\end{equation}
where
\begin{align}
 T^m&=
 m\left[
 3B\mu^2\left(m+(1+k^2)a^2kB\right)
 +a^2k\left(
 B^2\left(m^2+(1+k^2)^2a^2\right)+kmB-(1+k^2)
 \right)
 \right],
 \nonumber\\
 T^a&=
 a\left[
 B\mu^2\left(m+(1+k^2)a^2kB\right)
 -a^2k\left(
 B^2\left(m^2+(1+k^2)^2a^2\right)+kmB-(1+k^2)
 \right)
 \right],
 \nonumber\\
 T^e&=
 2eB\mu^2\left(m+(1+k^2)a^2kB\right),
 \nonumber\\
 T^B&= \left(m+(1+k^2)a^2kB\right)
 \left[1-B^2\left(2\mu^2+(2+k^2)a^2-e^2\right)\right]
 -a^2kB(\chi BK)^2.
\end{align}
Accordingly, the integrability equation can be decomposed into the constraint equation 
\begin{equation}
    V_i\cdot \delta \mathcal{A}=0,
\end{equation}
and the evolution equation
\begin{equation}
    T\cdot\delta \mathcal{A}=0
\end{equation}
by viewing $B$ as the time function,
where $V_i$ denote the three canonical coordinate tangential vector fields to the surface of constant $B$\footnote{The above constraint equation and the evolution equation are obviously equivalent to their restriction onto the surface of constant $B$.}. Note that $T\cdot\mathcal{A}=0$, the evolution equation is equivalent to 
\begin{equation}
 \mathcal L_T\mathcal{A}
 =T\cdot\delta\mathcal{A}+\delta(T\cdot\mathcal{A})
 =T\cdot\delta\mathcal{A}=0.
 \label{eq:moduli-Cartan}
\end{equation}
This implies that with our initial $\mathcal{A}$ (\ref{iv}), which apparently satisfies the constraint equation on $B=0$ surface, the push-forward of $\mathcal{A}$ by the diffeomorphism generated by $T$ gives rise to the solution to the integrability equation. To be more specific, according to 
\begin{equation}
 \mathcal L_T\nu^1=C_1\nu^1+C_2\nu^2+C_3\nu^3,
 \qquad
 \mathcal L_T\nu^2=\mathcal L_T\nu^3=0
 \label{eq:moduli-Lienu}
\end{equation}
with
\begin{align}
 &C_1=
 B((m+(1+k^2)a^2kB))
 [\mu^2+(2+k^2)a^2-e^2]
 +a^2k(\chi BK)^2
 ,
 \nonumber\\
 &C_2=
 eK(\chi BK)\left(1+k^2-\frac{e^2}{a^2}\right),
 \qquad\qquad\qquad
 C_3=\frac{me^2(\chi BK)}{a},
 \label{eq:moduli-Ccoefficients}
\end{align}
we have 
\begin{equation}
    \mathcal L_T\mathcal{A}={}
 (\mathcal{L}_T\gamma+C_1\gamma)\nu^1
 +[\mathcal{L}_T(\gamma\omega)+C_2\gamma]\nu^2
 +[\mathcal{L}_T(\gamma\alpha)+C_3\gamma]\nu^3.
 \label{eq:moduli-Liezeta}
\end{equation}
The linear independence of $\nu^i$, due to the fact that $T$ is non-vanishing, implies the following three evolution equations 
\begin{equation}
 \mathcal{L}_T\gamma=-C_1\gamma,
 \qquad
 \mathcal{L}_T(\gamma\omega)=-C_2\gamma,
 \qquad
 \mathcal{L}_T(\gamma\alpha)=-C_3\gamma,
 \label{eq:moduli-producttransport}
\end{equation}
which can be further cast into the three decoupled first-order ordinary differential equations 
\begin{equation}
 \mathcal{L}_T\alpha=C_1\alpha-C_3,  \quad  \mathcal{L}_T\gamma=-C_1\gamma,\quad
    \mathcal{L}_T\omega=C_1\omega-C_2
\end{equation}
along the integral curves of $T$. With the initial condition \eqref{initial}, the exact solution is given by
\begin{align} \alpha&=
 -\frac{e^2\left(\frac ea-k\chi BK\right)}{a(m(\chi BK)^2+e^2kB)},\nonumber\\
    \gamma&=\frac{m(\chi BK)^2+e^2kB}{\sqrt{K(
 m^2+e^2K-a^2k^2(1+k^2-\frac{e^2}{a^2}))
 }},
\nonumber\\
 \omega&=
 -\frac{
 aKB\left[
 k\left(1+k^2-\frac{e^2}{a^2}\right)
 +(1+k^2)mB
 \right]
 }{m(\chi BK)^2+e^2kB}.
 \label{appago}
\end{align}
Accordingly, one can further show $\mathcal A=\delta f$ with
\begin{equation}
f=
\frac{\sqrt{m^2+e^2(K+k^2)-a^2k^2(1+k^2)}}{K^{3/2}}.
\end{equation}

\end{document}